\documentclass[osajnl,twocolumn,showpacs,superscriptaddress,10pt]{revtex4-1} 

\usepackage{graphicx} \usepackage{amsmath, amssymb, bm}
\usepackage{dcolumn} \usepackage{color}
\usepackage{eso-pic}
\usepackage{fix-cm}

\usepackage[capbesideposition=right]{floatrow}
\usepackage{multirow}
\usepackage{sidecap}

\usepackage{marginnote}
\usepackage{ulem} 

\newcommand{\sss}{\scriptscriptstyle}
\newcommand{\sst}{\scriptstyle}

\newcommand{\stext}[1]{\sss \text{#1} \sst}

\renewcommand{\emph}[1]{\textit{#1}}

\begin{document}

\title{Cavity-enhanced optical Hall effect in two-dimensional free charge carrier gases detected at terahertz frequencies }

\author{S. Knight}
\affiliation{Department of Electrical and Computer Engineering and Center for Nanohybrid Functional Materials, University of Nebraska-Lincoln, Lincoln,
Nebraska, 68588-0511, USA}
\author{S. Sch\"{o}che}
\affiliation{J.A. Woollam Co., Inc., 645 M Street, Suite 102, Lincoln, Nebraska 68508-2243, USA}
\author{V. Darakchieva}
\affiliation{Department of Physics, Chemistry, and Biology (IFM), Link\"{o}ping University, SE 581 83 Link\"{o}ping, Sweden}
\author{P. K\"{u}hne}
\affiliation{Department of Physics, Chemistry, and Biology (IFM), Link\"{o}ping University, SE 581 83 Link\"{o}ping, Sweden}
\author{J.-F. Carlin}
\affiliation{Ecole Polytechnique F\'{e}d\'{e}rale de Lausanne (EPFL), 1015 Lausanne, Switzerland}
\author{N. Grandjean}
\affiliation{Ecole Polytechnique F\'{e}d\'{e}rale de Lausanne (EPFL), 1015 Lausanne, Switzerland}
\author{C.M. Herzinger}
\affiliation{J.A. Woollam Co., Inc., 645 M Street, Suite 102, Lincoln, Nebraska 68508-2243, USA}
\author{M. Schubert}
\affiliation{Department of Electrical and Computer Engineering and Center for Nanohybrid Functional Materials, University of Nebraska-Lincoln, Lincoln,
Nebraska, 68588-0511, USA}
\author{T. Hofmann}
\affiliation{Department of Electrical and Computer Engineering and Center for Nanohybrid Functional Materials, University of Nebraska-Lincoln, Lincoln,
Nebraska, 68588-0511, USA}
\affiliation{Department of Physics, Chemistry, and Biology (IFM), Link\"{o}ping University, SE 581 83 Link\"{o}ping, Sweden}
\homepage{http://ellipsometry.unl.edu}

\date{\today}
\email{Corresponding author: thofmann@engr.unl.edu}

\begin{abstract}The effect of a tunable, externally coupled Fabry-P\'{e}rot cavity to resonantly enhance the optical Hall effect signatures at terahertz frequencies produced by a traditional Drude-like two-dimensional electron gas is shown and discussed in this communication. As a result, the detection of optical Hall effect signatures at conveniently obtainable magnetic fields, for example by neodymium permanent magnets, is demonstrated. An AlInN/GaN-based high electron mobility transistor structure grown on a sapphire substrate is used for the experiment. The optical Hall effect signatures and their dispersions, which are governed by the frequency and the reflectance minima and maxima of the externally coupled Fabry-P\'{e}rot cavity, are presented and discussed. Tuning the externally coupled Fabry-P\'{e}rot cavity strongly modifies the optical Hall effect signatures, which provides a new degree of freedom for optical Hall effect experiments in addition to frequency, angle of incidence and magnetic field direction and strength.
\end{abstract}



\maketitle

The optical Hall effect (OHE) in semiconductor layer structures is the occurrence of magneto-optic birefringence detected in response to incident electromagnetic radiation, caused by movement of free charge carriers under the magnetic field-induced influence of the Lorentz force~\cite{SchubertJOSAA20_2003}. In general, this birefringence leads to polarization mode coupling which is conveniently detected by generalized ellipsometry at oblique angle of incidence and at terahertz (THz) frequencies, for example. THz-OHE has recently been demonstrated as non-contact and therefore valuable tool for the investigation of free charge carrier properties in semiconductor heterostructures
\cite{KuehneRSI85_2014, KuehnePRL111_2013, HofmannAPL101_2012, HofmannTSF519_2011, HofmannAPL98_2011, SchocheAPL98_2011,Hofmannpss205_2008, HofmannRSI77_2006}. Previous instrumental approaches, discussed more detailed in Refs.~\onlinecite{HofmannRSI77_2006, KuehneRSI85_2014}, rely on high magnetic fields provided either by conventional, water-cooled or superconducting, liquid He-cooled electromagnets resulting in comparably large and costly experimental setups. In general, OHE configurations capable of detecting signals at low and conveniently obtainable magnetic fields are desirable. The use of small magnetic fields for THz magneto-optic measurements was demonstrated recently by Ino \emph{et al.} for bulk-like InAs \cite{InoPRB70_2004}. Due to low effective mass and high charge carrier concentration, the low field still yielded large enough signals for detection. The signal-to-noise separations of the OHE signatures depend on many factors, the most important are low effective mass, high mobility and high carrier density, but also crucial is the thickness of the physical layer that contains the charge carriers. The OHE signatures scale, in first approximation, linearly with the magnetic field amplitude. Hence, the first approach to detect OHE signatures in samples with very thin layers, or low-mobile, heavy-mass and low-density charge carriers where the OHE signatures are weak is to increase the magnetic field amplitude.

\begin{figure}
\centering
\includegraphics[keepaspectratio=true,width=7.5cm, clip, trim=0 00 0 0 ]{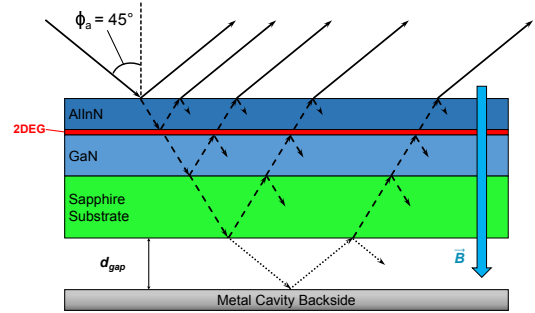}
\caption{Schematic drawing of the beam path through the sample and the external optical cavity, shown for example for an AlInN/GaN/sapphire high electron mobility transistor (HEMT) structure with two-dimensional electron gas (2DEG). The sapphire substrate and metallic cavity surface are parallel and separated by the distance $d_{\stext{gap}}$. Here, the magnetic field $\mathbf{B}$ is perpendicular to the sample surface with the positive magnetic field direction oriented into the sample. The metal cavity backside may be formed by an appropriately surface-coated permanent magnet. Note that HEMT layer structure, substrate and cavity are not to scale.\vspace{-0.5cm}}
\label{fig:fig_principle}
\end{figure}

In this communication, we demonstrate and exploit the enhancement of the OHE signal obtained from samples with plane parallel interfaces deposited on THz transparent substrates using an external and tunable optical cavity. We show that an OHE signal enhancement of up to one order of magnitude can be achieved by optimizing the cavity geometry, which is very useful for small magnetic field strengths. This signal enhancement allows the determination of free charge carrier effective mass, mobility, and density parameters using OHE measurements at low magnetic fields. The OHE signals are defined by and presented here as the differences between the Mueller matrix elements determined for opposing magnetic field directions~\cite{KuehneRSI85_2014}. An AlInN/GaN-based high electron mobility transistor structure (HEMT) grown on a sapphire substrate is investigated as an example, while the cavity enhancement phenomenon discussed here is generally applicable to situations when a layered sample is deposited onto a transparent substrate. This cavity enhancement may be exploited in particular for layered samples grown on technologically relevant low-doped or semi-insulating substrate materials such as SiC, Si, or GaAs, etc.

\begin{figure*}
\centering
\caption{Model-calculated contour plots of typical THz-OHE data, here for example $\Delta M_{13,31} = M_{13,31}(+B) - M_{13,31}(-B)$ and $\Delta M_{23,32} = M_{23,32}(+B) - M_{23,32}(-B)$ for the AlInN/GaN HEMT sample are shown as a function of frequency and $d_{gap}$. The vertical solid and dashed black lines indicate the sample's $s$-polarized reflection maxima and minima, respectively. The $s$-polarized reflectivity maxima and minima of the external cavity and which depend on $d_{\stext{gap}}$, are shown as horizontal solid and dashed black lines, respectively. All data was calculated for an angle of incidence $\Phi_{a}$=45$^{\circ}$ and a magnetic field magnitude of $|B|=0.55~$T. Note that the $p$-polarized modes occur slightly shifted with respect to the $s$-polarized modes and are omitted for clarity.\vspace{-0.5cm}}\label{fig:calc}
\includegraphics[keepaspectratio=true,width=18cm, clip, trim=10 0 0 0 ]{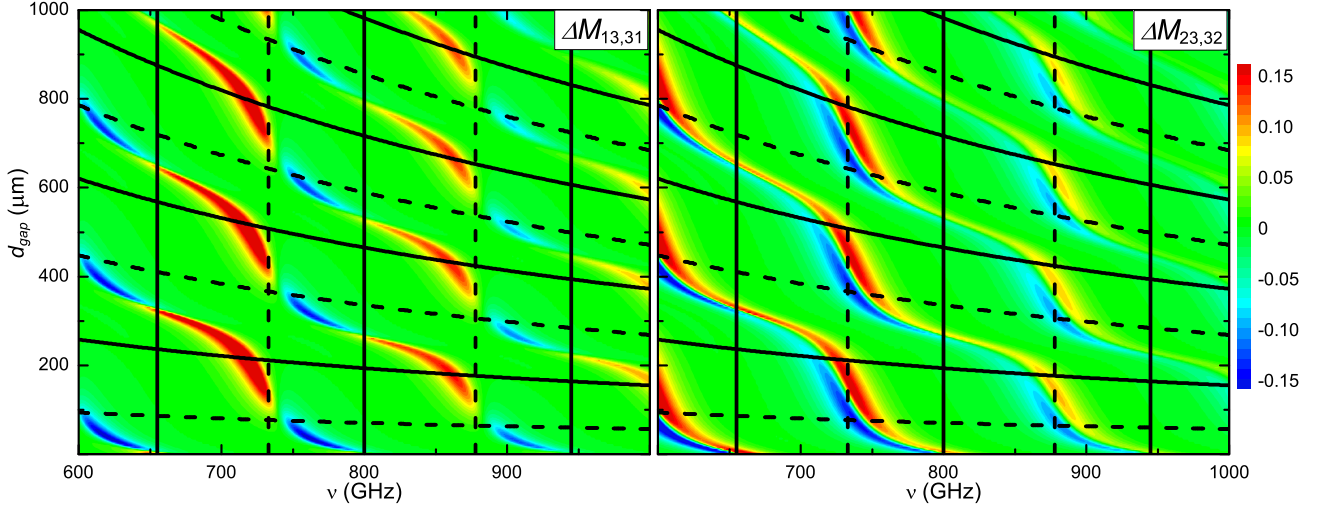}
\end{figure*}

For a thin-film layer stack deposited on a transparent substrate, where the substrate thickness is much larger than the combined layer thickness of all sublayers in the stack, and where the substrate thickness may be multiple orders of the wavelength at which the OHE signatures are detected, the fraction of the incident beam transmitted through the entire sample is coupled back into the substrate using an external cavity. For example, a highly reflective surface placed at a distance $d_{\stext{gap}}$ behind and parallel to the backside of the substrate, as shown in Fig.~\ref{fig:fig_principle}, permits THz radiation to be coupled back into the sample and thereby produce an enhancement of the OHE signal. In our example discussed below we have achieved up to one order of magnitude enhancement by varying $d_{\stext{gap}}$.

Figure~\ref{fig:calc}, shows model-calculated contour plots of the THz-OHE signal (difference of the Mueller matrix elements calculated for $B=0.55$~T and $B=-0.55$~T) illustrating the enhancement phenomenon. The structure used for the calculation is a AlInN/GaN HEMT structure deposited on a 350~$\mu$m thick $c$-plane Al$_2$O$_3$ substrate, similar to the HEMT structure discussed in Ref.~\onlinecite{SchocheAPL98_2011}. The non-trivial off-block Mueller matrix components $\Delta M_{13,31}$ and $\Delta M_{23,32}$ show periodic resonances which depend on the frequency of the THz probe beam $\nu$ and $d_{\stext{gap}}$. The frequency is varied over the range from 600 to 1000~GHz and the $d_{\stext{gap}}$ ranges from 0 to 1000~$\mu$m to obtain overview over an experimentally feasible parameter range and to gain insight into the multiplicity of the occurrences of coupled-substrate-cavity mode enhancements of the OHE signal.
In order to show that these occurrences are related to the minima in reflectance for the substrate and cavity modes, the maxima (minima) of the $s$-polarized reflectivity of the sample and the cavity are plotted as solid (dashed), vertical and horizontal lines, respectively. It can be seen that the largest OHE signal is expected in the vicinity of the intersections of the reflection extrema (maxima or minima) of sample and cavity Fabry-P\'{e}rot modes. The resonance frequency of the sample Fabry-P\'{e}rot mode is determined by the sample's substrate thickness which is much larger than the total HEMT thickness (see further below). The $p$-polarized modes which occur slightly shifted with respect to the $s$-polarized modes are omitted for clarity in Fig.~\ref{fig:calc}.
It is interesting to note, that there are regions in frequency and $d_{\stext{gap}}$ where the OHE signal is very small or vanishes. Hence, experimental configurations where both frequency and $d_{\stext{gap}}$ can be varied over sufficiently large regions, that is, to cover at least one period of coupled substrate-cavity modes will be valuable for practical applications.

\begin{figure*}
\fcapside[13.0cm]
{\caption{Panels a) and b) show the same contour plots as in Fig.~\ref{fig:calc} for the frequency range and gap distances investigated by experiment here.
The panels c) and d) show the corresponding experimental (green dashed lines) and best-model calculated (red solid lines) data $\Delta M_{13,31}$ and $\Delta M_{23,32}$  at three different $d_{\stext{gap}}$ values indicated as horizontal black lines in the a) and b).
The panels c) and d) also include best-model calculated data for $d_{\stext{gap}} \rightarrow \infty$  as blue solid lines for comparison. All data are obtained at room temperature at an angle of incidence $\Phi_{a}$=45$^{\circ}$ and a magnetic field magnitude of $|B|=0.55~$T.}\label{fig:exp}}
{\includegraphics[keepaspectratio=true,width=12.0cm, clip, trim=00 0 5 10 ]{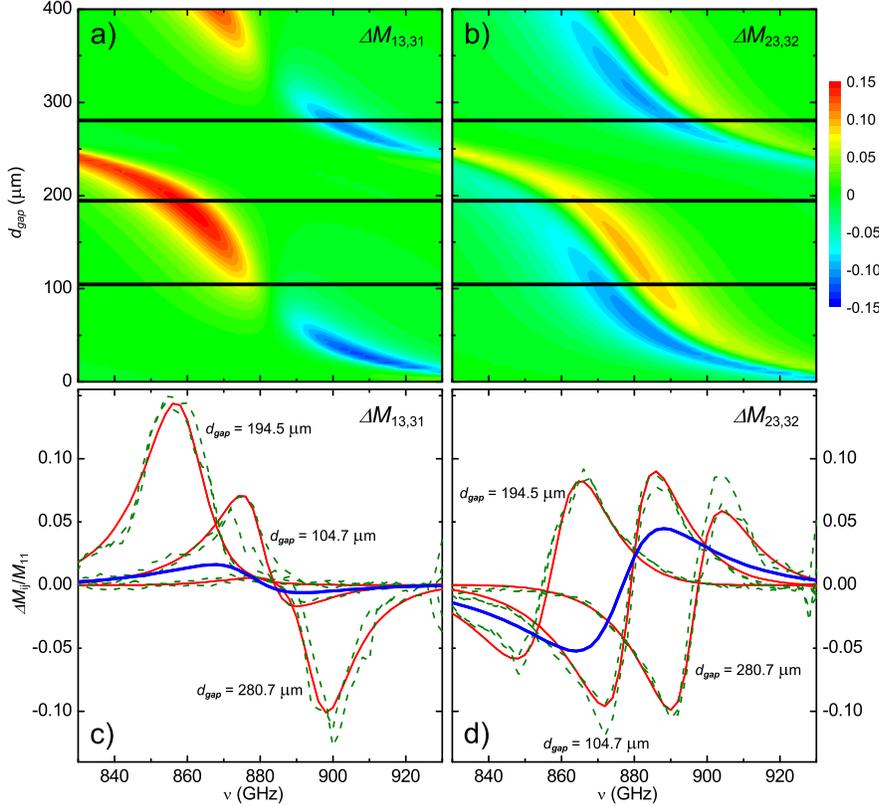}
}\vspace{-0.3cm}
\end{figure*}

For the experimental verification of this enhancement effect an AlInN/GaN-based HEMT structure was grown using metal-organic vapor phase epitaxy on a single side polished $c$-plane sapphire substrate with a nominal thickness of 350~$\mu$m. Subsequent to the growth of a 2~$\mu$m  thick undoped GaN buffer layer, a 1~nm thick AlN spacer layer was deposited, followed by a 12.3~nm thick Al$_{0.82}$In$_{0.18}$N top layer \cite{DarakchievaJAP08}. The THz-OHE data presented here were obtained using a custom-built THz ellipsometer \cite{HofmannRSI81_2010,KuehneRSI85_2014}. THz-OHE data were measured in the spectral range from 830 to 930~GHz with a resolution of 2~GHz at an angle of incidence $\Phi_a=45^{\circ}$ and for three different gap distances $d_{\stext{gap}}$ of 104.7~$\mu$m, 194.5~$\mu$m, and 280.7~$\mu$m to cover at least one of the coupled substrate-cavity mode features rendered in Fig.~\ref{fig:exp}.

The measurements were facilitated by mounting the HEMT structure onto a Ni-coated, high-grade N42 neodymium permanent magnet using adhesive spacers to create a homogeneous air gap between the Ni-coated surface of the magnet which serves as the metallic cavity backside and the HEMT structure. The THz Mueller matrix measurements  were carried out with the sample mounted on the north and on the south pole-face of the permanent magnet to obtain THz-OHE data (differences of the Mueller matrix elements $M_{13}$, $M_{23}$, $M_{31}$, and $M_{32}$ measured at opposing magnetic fields). Across the sample area illuminated by the THz probe beam, the magnetic field strength provided by the permanent magnet was $B$ = ($0.55\pm0.005$)~T determined using a Hall sensor (Lakeshore). For values of $d_{\stext{gap}}$ used here, the change in the magnetic field magnitude at different gap values is negligible at the sample position.

In addition to the THz-OHE measurements, the sample as well as the metal magnet surfaces were investigated using a commercial (J.A. Woollam Co. Inc.) mid-infrared (MIR) ellipsometer in the spectral range from 300 to 1200~cm$^{-1}$ at $\Phi_a$~=~60$^{\circ}$ and 70$^{\circ}$ in order to determine the HEMT layer thickness parameters and phonon mode parameters, and the optical constants of the magnet surface metal layer (Ni). All measurements were carried out at room temperature and analyzed simultaneously. The experimental and model calculated data are reported using the Mueller matrix formalism \cite{Fujiwara_2007}.

The experimental MIR-SE and THz-OHE data sets were analyzed simultaneously using an optical model composed of eight phases including a AlInN top layer/2DEG/AlN spacer/GaN buffer/Al$_2$O$_3$ substrate/air gap/Ni cavity surface~\cite{SchocheAPL98_2011}. Nonlinear regression methods where used to match the lineshape of experimental and optical model calculated data as close as possible by varying relevant model parameters using parameterized model dielectric functions \cite{Fujiwara_2007}. The THz and MIR dielectric function tensors of the optically uniaxial sample constituents GaN, AlInN, AlN and Al$_2$O$_3$ are composed of contributions from optically active phonon modes $\varepsilon^{\text{L}}(\omega)$ and free-charge carrier excitations $\varepsilon^{\text{FC}}(\omega)$. Details on the parametrization approach are omitted here for brevity and we refer to previous publications \cite{SchubertJOSAA20_2003, SchubertIRSEBook_2004,Pidgeon80,Hofmannpss205_2008,Yu99,Pidgeon80}.
The optical response of the magnet's Ni mirror surface that forms the external cavity is described using the classical Drude formalism using the static resistivity parameter of $\rho = (1.72\pm0.49)~\times~10^{-5}~\Omega$cm and the average-collision time parameter of $\tau = 6.15~\times~10^{-16}~$s. $\rho$ is obtained as best-match model parameter from MIR-SE data analysis, and is within typical values for Ni \cite{AbdallahTSF571_2014,RobertsPRL114_1959}. The average-collision time parameter is taken from Ref.~\onlinecite{AbdallahTSF571_2014} and not varied in the model analysis. 

The top panels of Fig.~\ref{fig:exp} enlarge the contour plots from Fig.~\ref{fig:calc} for frequencies and external cavity distances at which OHE experiments were performed. The horizontal lines indicate the fixed gap values used here. The bottom panels of Fig.~\ref{fig:exp} show experimental (green dashed lines) and best-model calculated (red solid lines) THz-OHE spectra (differences of the Mueller matrix elements $M_{13}$, $M_{23}$, $M_{31}$, and $M_{32}$) measured at $B=0.55$~T and $-0.55$~T. For comparison the bottom panels of Fig.~\ref{fig:exp} also include best-model calculated data for $d_{\stext{gap}} \rightarrow \infty$  as blue solid lines. The amplitude of $\Delta M_{13}$, $\Delta M_{31}$, $\Delta M_{23}$, and $\Delta M_{32}$ is proportional to the percentage of cross-polarization at the given frequency \cite{Hofmannpss205_2008,SchubertIRSEBook_2004} and is caused only by the Drude-like magneto-optic contribution of the 2DEG. Based on the best-model analysis the low background free charge carrier densities of the AlInN, GaN, and AlN layers were found to have a negligible contribution to the THz-OHE signal. We find a good agreement between experimental and best-model calculated data for the different $d_{\stext{gap}}$ values.

The maxima and minima depicted in the best-model calculated contour plots of $\Delta M_{13,31}$ and $\Delta M_{23,32}$ (top panels of Fig.~\ref{fig:exp}) are due to the coupling of the Fabry-P\'{e}rot oscillations in the sample structure with those of the external cavity.
The experimentally accessed range in frequency and $d_{\stext{gap}}$ was selected to sufficiently cover the response of the HEMT sample-substrate-cavity mode under the influence of a small magnetic field to detect the enhanced OHE signal. Depending on the distance $d_{\stext{gap}}$ between sample backside and cavity surface the frequency dependent response of the OHE signal changes where extrema occur in the vicinity of the intersection of the sample and cavity reflection extrema. Comparing $\Delta M_{13,31}$ and $\Delta M_{23,32}$ reveals distinct differences. Whereas $\Delta M_{23,32}$ shows a derivative-like shape with a single pair of maximum and minimum in the range from 830 to 930~GHz, $\Delta M_{13,31}$ exhibits a single maximum or minimum and a strong amplitude variation, depending on $d_{\stext{gap}}$. The largest amplitude is observed for $d_{\stext{gap}}=194.5~\mu$m where $\Delta M_{13,31}$ is approximately 0.15. The smallest change is observed for $d_{\stext{gap}}=104.7~\mu$m where
$\Delta M_{13,31}\approx0.07$. The best-model calculated data excluding the cavity enhancement shown as solid blue line in Fig.~\ref{fig:exp} c) is almost vanishing and the cavity enhances the OHE signal by one order of magnitude. The largest amplitudes of $\Delta M_{23,32}\approx 0.1$ can be observed for $d_{\stext{gap}}=104.7~\mu$m and 280.7$~\mu$m (Fig.~\ref{fig:exp} d). The largest amplitudes in $\Delta M_{23,32}$ without the cavity effect is approximately 0.05 which is a factor of two smaller than the OHE signal amplitude observed for $d_{\stext{gap}}=104.7~\mu$m and 280.7$~\mu$m.

The best-model sheet density, mobility, and effective mass obtained for the 2DEG are $N = (1.02\pm0.15) \times 10^{13}~\text{cm}^{-2}$, $\mu = (1417\pm97)~\text{cm}^{2}/\text{Vs}$, $m^{\ast} = (0.244\pm0.020) m_{0}$, respectively. These results are in good agreement with the results of  high-field ($B=7$~T) THz-OHE measurements on the same sample $N = (1.40\pm0.07) \times 10^{13}~\text{cm}^{-2}$, $\mu = (1230\pm36)~\text{cm}^{2}/\text{Vs}$, $m^{\ast} = (0.258\pm0.005)\ m_{0}$. Our findings demonstrate that the cavity enhancement of the THz-OHE signal allows the investigation of free charge carrier properties of two dimensional free charge carrier gases at low magnetic fields which may be conveniently provided by permanent magnets.

A variation in $d_{\stext{gap}}$ provides a new degree of freedom to tune the experimental conditions so as to reach a maximum response for a given frequency range. According to the contour plot in Fig.~\ref{fig:exp} c) and d), other gap lengths, e.g., 150~$\mu$m will provide even larger signals in this situation. For a given sample system the experimental configuration can be optimized by calculating the coupled substrate-cavity modes and then selecting frequency range and $d_{\stext{gap}}$ accordingly. Furthermore, varying the $d_{\stext{gap}}$ at a fixed frequency may be used to maximize the OHE signal which can be very useful for single-frequency GHz or THz instrumentation.

In this letter, we demonstrate that the THz-OHE signal for samples grown on THz transparent substrates can be controlled and enhanced by a tunable, externally coupled Fabry-P\'{e}rot cavity mode. In the vicinity of a coupled mode, a strong enhancement of the OHE signatures is observed. Using the optical cavity enhancement THz-OHE measurements at small magnetic fields provided by a permanent magnet can be detected. We demonstrate this effect here by determination of the free charge carrier properties of a high electron mobility transistor structure using a permanent magnet. Previously, superconducting high-field electromagnets needed to be employed for this determination. For our example, tuning the external cavity allows an enhancement of the OHE signatures by as much as one order of magnitude. We propose to employ this enhancement effect to reliably and accurately determine free charge carrier properties in layered semiconductor samples at small magnetic fields dispensing with the need for expensive high-field electromagnets.

The authors acknowledge financial support from the National Science Foundation through the Nebraska Materials Research Science and Engineering Center (MRSEC) (grant No. DMR-1420645) and through the Center of Nanohybrid Materials (grant No. EPS-1004094), the J.A. Woollam Foundation, and the University of Nebraska-Lincoln. We furthermore acknowledge support from the Swedish Research Council (grant No. 2013-5580), the Swedish Agency for Innovation Systems (grants No. 2011-03486 and No. 2014-04712), and Swedish Foundation for Strategic Research (grant No. FFL12-0181).


\end{document}